# An Integrated Experimental and Modeling Approach for Crystallization of Complex Biotherapeutics

Vivekananda Bal[1], Moo Sun Hong,[2] Jacqueline M. Wolfrum[3], Paul W. Barone[3], Stacy L. Springs[3], Anthony J. Sinskey[3,4], Robert M. Kotin[3,5], and Richard D. Braatz[1,3]

[1] Department of Chemical Engineering, Massachusetts Institute of Technology, Cambridge, MA, USA

[2] School of Chemical and Biological Engineering, Seoul National University, Seoul, Republic of Korea

[3] Center for Biomedical Innovation, Massachusetts Institute of Technology, Cambridge, MA, USA

[4] Department of Biology, Massachusetts Institute of Technology, Cambridge, MA, USA

[5] Gene Therapy Center, University of Massachusetts Chan Medical School, Worcester, MA, USA

**Abstract:** Crystallization of proteins, specifically proteins of medical relevance, is performed for various reasons such as to understand the protein structure and to design therapies. Obtaining kinetic constants in rate laws for nucleation and growth of advanced biotherapeutics such as capsids, an assembly of macromolecules, is challenging and essential to the design of the crystallization processes. In this work, coupled population balance and species balance equations are developed to extract nucleation and growth kinetics for crystallization of recombinant adeno-associated virus (rAAV) capsids. A comparison of model results with that of experimental data for capsid crystallization in hanging-drop vapor diffusion system shows that slow rate of vapor diffusion from the droplet controls the initial nucleation and growth processes, and the capsid nucleation occurs via heterogeneous nucleation in the microdroplet. Results also show that the capsids, which are of very high molecular weight (~3.6 MDa), have a similar tendency to nucleate as small organic molecules such as glycine (75 Da), low-molecular-weight proteins, and small-molecule active pharmaceutical ingredients due to its ball-shaped outer structure/shape. Capids also show a prolonged nucleation period as for proteins and other macromolecules, but has a slow growth rate with a growth rate pre-factor seven orders of magnitude smaller than that of lysozyme. The capsid crystal growth rate is weakly sensitive to the supersaturation compared to lysozyme and is limited by the transport of capsids due to slow Brownian motion resulting from the very high molecular weight.

## 1. Introduction

Protein crystallization is important in the field of bioscience and biotechnology research as it has numerous applications including protein purification, low-viscosity drug substance and drug product formulation, drug discovery, and target identification and selection.[1,2] Apart from this, the knowledge of the detailed 3D structure of proteins facilitates the understanding of many fundamental mechanisms involved in biological processes such as how interactions among biological macromolecules helps in building supramolecular nano-machines able to perform specific biological functions.[3,4] X-ray diffraction, which determines the 3D structure of protein from crystals, needs higher quality/diffraction-grade crystals.[1,2] However, growing diffraction-grade crystals can be extremely difficult as suggested by a comparison between the number of non-redundant protein sequences deposited in Uniprot (> 20 million) and protein structures deposited in the Protein Data Bank (about 196,400 as of April 2024).[2]

Hanging-drop vapor diffusion experiment is widely used in protein crystallization to grow diffraction-grade crystals or for analytical screening purposes due to requiring a low quantity of material, simple operation, and easy monitoring of crystals.[1,2,5–10] Growing diffraction-grade crystals requires in-depth understanding of the system and the crystallization process, the influence of solution parameters, knowledge of crystallization kinetics, and control of crystallization conditions. An integrated experimental and modeling approach can be useful in this regard as they are complementary. Though mathematical modeling of crystallization processes for small-molecule pharmaceuticals is common,[11–15] and for proteins such as lysozyme (~14.3 kDa) have been explored to some extent,[16–18] there are no descriptions of modeling of the crystallization process for large biological macromolecules such as monoclonal antibodies (mAbs, ~150 kDa), mRNA (200–500 kDa), and complex proteinaceous assemblies, such as viral capsids (MW ~ 3.6 MDa).

There are some mathematical models describing the crystallization of small proteins such as lysozyme, a globular protein of 14.3 kDa, but those models are mainly applicable for relatively large-scale processes and/or do not consider the mechanistic details.[4,7,16,18] Some of these studies propose mechanisms of polynuclear crystal growth and self-assembly nucleation events while other reports follow classical models. As far as crystallization in hanging-drop vapor diffusion system is concerned, there are no reports describing the crystallization of proteins/proteinaceous assemblies/macromolecular assemblies using integrated experimental and population balance modeling approaches that incorporate mechanistic details and solution thermodynamics.

In this work, we develop a mathematical model based on coupling population balance and species balance equations for hanging-drop vapor diffusion system. The modelling includes the effects of droplet volume reduction and solution thermodynamics that are validated with experimental data for capsid crystallization. This work uses recombinant adeno-associated virus (rAAV) capsids as a model biological molecule. rAAV is a commonly used vector for gene transfer/therapy. The icosahedral virus particle, aka capsid, is assembled from 60 protein subunits, VP1 (87 kDa), VP2 (72 kDa), and VP3 (62 kDa) in roughly a 1:1:10 ratio.[19] AAVs are available in different serotypes 1 through 11. Two serotypes of rAAV, derived from AAV5 and AAV8, were selected for our study. **Fig. 1a** shows the corresponding capsid models, where capsids are radially colored by distance from the center (obtained from VIPER database,[20] https://viperdb.org) indicating the location of proteins and showing some depression and protrusion regions. These two different AAV species are about 58% identical across VP1 and the same percent identity between the major capsid proteins, VP3.[21] **Fig. 1b** shows the capsid surface with different VP3 subunits marked by different colors.[22] Each subunit is colored such that subunits do not contact another subunit of the same color. From the models of intact capsids as shown in **Fig. 1a**, it is difficult to understand the slight differences between two serotypes. The best way to visualize the differences between the two serotypes is by aligning VP3 ribbon or mesh surface models as shown in **Fig. 1c**.

Recombinant AAV manufacturing processes result in a heterogenous AAV vector product containing "full" capsids, which contain singlestranded linear DNA (referred to as the vector genome, vg) and are biologically active, and "empty" capsids, which lack the genome and do not contribute to the biotherapeutic activity and therefore may be considered as a process impurity. Separating the empty from the full capsids is difficult, however, due to the similar physicochemical properties of the two particles. For full and empty capsids, the calculated pI for full and empty capsids, are 5.9 and 6.3, molecular weights are 5.8 MDa and 3.8 MDa, and densities are $1.41 gm/cm^3$ and $1.31 gm/cm^3$, respectively. Thus, understanding their crystallization kinetics will be useful in designing a separation process based on crystallization. Though this work uses rAAV capsids, a macromolecular assembly, as a model biological molecule, the mathematical modelling framework can also be applied for understanding the crystallization process and predicting solution conditions suitable for crystallization of other proteins including advanced therapeutics, small molecules, and macromolecules. This modelling framework is also applicable to other droplet-based crystallization systems such as sitting- and sandwich-drop vapor diffusion systems.

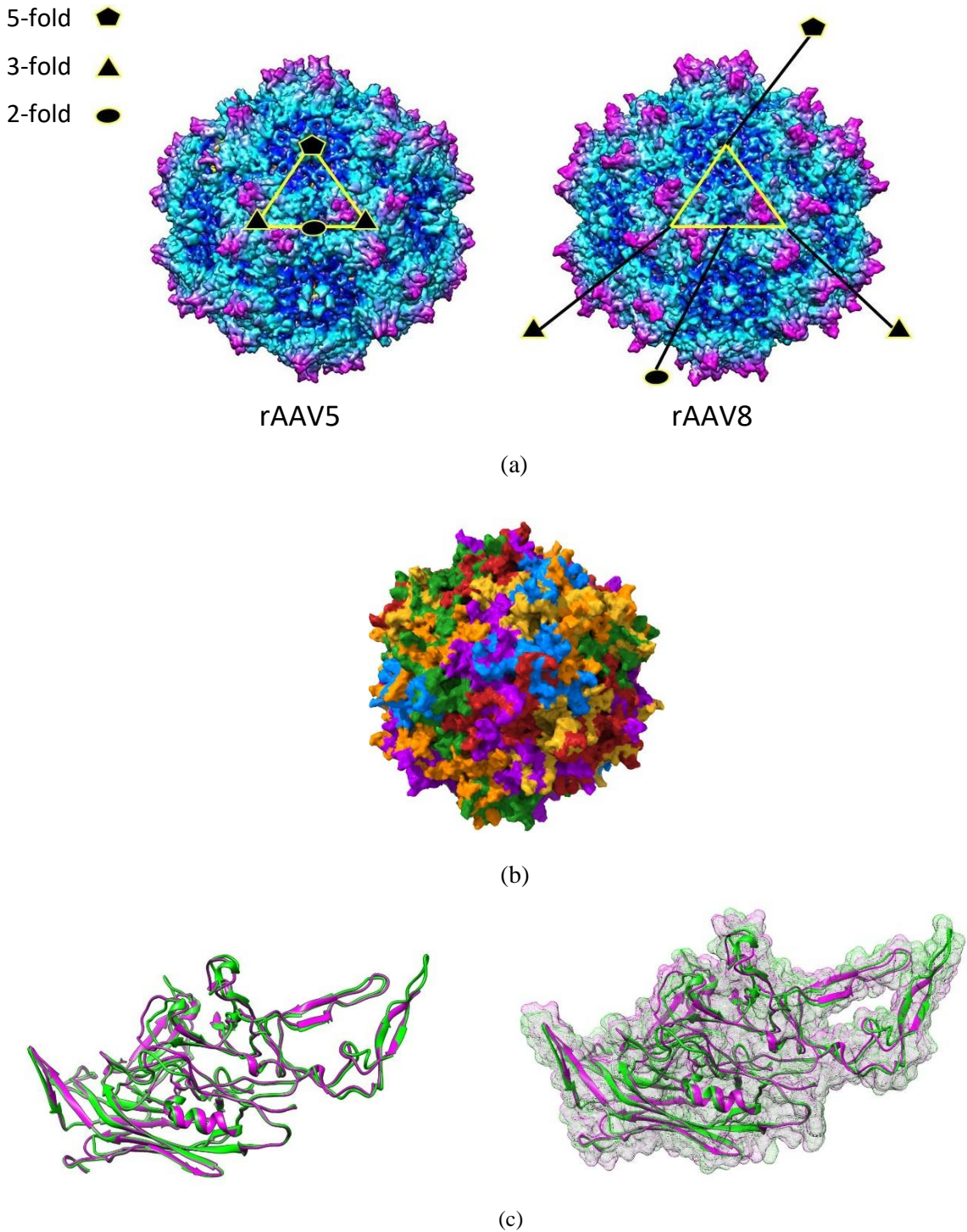

**Fig. 1:** (a) AAV5 and AAV8 capsid models produced using UCSF Chimera. The capsids are rendered at 2 nm resolution and are radially colored by distance from the center (red 9 nm, yellow, 10 nm, dark blue 11 nm, cyan 12 nm, violet 14 nm). The atomic coordinates were obtained from the VIPER database[20] (https://viperdb.org). Models showing the 2-, 3-, and 5- fold symmetry axis for capsid as indicated by the circle, triangle, and pentagon, respectively (b) The capsid surface showing different VP3 subunits. Each VP3 subunit is colored such that the subunits do not contact another subunit of the same color.[22] (c) The ribbon (left) and mesh (right) surface models of the VP3 subunits of AAV5 (magenta) and AAV8 (green) capsids.

### 2.1 Materials
Chemicals used in the experiments were of molecular biology grade. Sodium dihydrogen phosphate dihydrate (BioUltra, ≥99.0%), 1M hydrochloric acid (BioReagent, for cell culture), sodium hydroxide (BioXtra, ≥98% anhydrous), sodium chloride (BioXtra, ≥99.5%), potassium dihydrogen phosphate (BioUltra, ≥99.5%), potassium chloride (BioXtra, ≥99.5%), polyethylene glycol (PEG-8000, BioUltra; PEG-6000, BioUltra), and phosphate buffer saline (PBS 1X (150 mM sodium phosphate and 150 mM NaCl), pH 7.2; BioUltra solution) were purchased from the Sigma-Aldrich. Poloxamer-188 (Pluronic F-68, 10%, BioReagent) was purchased from Thermo Fisher and added to rAAV-containing solutions at 0.001% to suppress binding to container surfaces.

### 2.2 AAV samples
Full capsids (capsid carrying transgene) of the recombinant adeno-associated virus serotypes 5 (rAA5) and 8 (rAAV8) were purchased from Virovek at a concentration of $10^{14}$ vg/ml and pH 7.2 in PBS buffer. The "full rAAV5" and "full rAAV8" samples actually contain 80% full and 20% empty capsids. The full capsids contained a transgene of length 2.5 kbp. Virovek deliver samples in vials, each carrying 100 µL of sample. The sample from each vial was divided into four equal parts and stored for long-term use at −80°C in a freezer. For immediate use, a small vial was stored at 4°C, at which AAV is stable for 4 weeks.

### 2.3 Experiment
All the experiments in this work are performed at room temperature, 23±2°C. Crystallizations were performed using the hanging-drop vapor-diffusion method (**Fig. 2**) in VDX 24-well crystallization plates with glass cover slips at the top (Hampton Research, California, USA). Each well contains a droplet (2 µL) suspended from the glass cover slip covering the well and a reservoir of volume 1 mL at the bottom of the well. Each reservoir solution contains polyethylene glycol (PEG) 8000 and sodium chloride (NaCl) dissolved in a phosphate-buffered saline (PBS) solution. The droplet contains 1 µL of rAAV sample and 1 µL of reservoir solution. The concentrations of PEG and salt in the droplet are half of that in the reservoir solution.

Over time, the water from the droplet diffuses into the volume of the well and equilibrates with the reservoir solution. As the vapor pressure of water in the droplet is greater than that in the reservoir, due to the relatively low salt concentration in the droplet (vapor pressure depression, colligative property), the droplet volume shrinks. After supersaturation is reached, the nucleation of AAV capsid crystals begins. Eventually, as the ionic strengths of the two solutions equilibrate, the vapor diffusion rate decreases and no further reduction in droplet volume occurs.

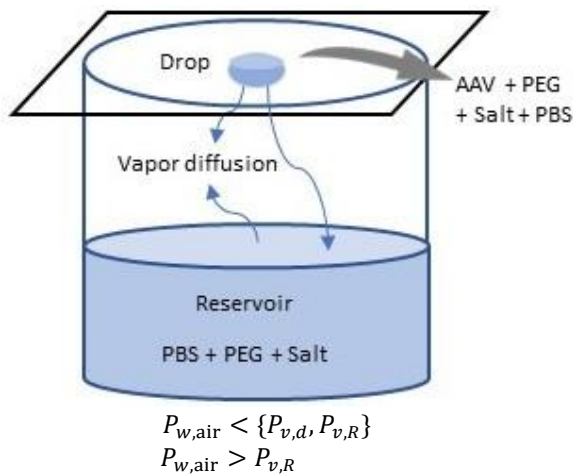

$P_{w,\text{air}} < \{P_{v,d}, P_{v,R}\}$
$P_{w,\text{air}} > P_{v,R}$

**Fig. 2:** Schematic of hanging-drop vapor diffusion experiment in one of the 24 wells in a VWR 24-well plate.

Each droplet was monitored at regular intervals for a period of 1 to 2 weeks via an optical microscope (Imaging Source DMK42BUC03) and a cross-polarized light miscroscope (Leica Z16 APO) to observe the outcome of the experiment. Images of the droplet containing crystals were captured using a high-resolution camera installed in an automated vapor diffusion setup at specific time intervals (see Ref. [7] for a detailed description on the automated setup). The 24-well VWR crystallization plate was placed under the high-resolution camera in the automated setup, images were captured, and the plate was removed from the instrument. Image J software was used to measure the crystal dimensions such as length and radius in each droplet image as obtained from the high resolution camera. Then the crystals were classified into different size classes. Each size class is a bin of specific size range containing crystals of sizes falling that range. Crystal surface area and volume was calculated from the crystal radius and length as obtained from the image J analysis. Total surface area of all the crystals in the droplet was calculated by multiplying the number of particles in each size class with the surface area of a crystal in that size class and then summing over all the size classes. The total volume of all the crystals was similarly calculated.

Some crystallization conditions produce a large number of crystals (~3000–4000) with different undefined morphologies. This work considers only those experimental conditions, which produce only a few crystals (~ 30–300) with well defined morphology such as needle, rod, and cylinder for the convenience of measurement.

### 3. Model Development

The design of a production-scale process for crystallization of full capsids requires characterization of the crystallization kinetics. Direct measurement of crystal nucleation rates is challenging due to the very small size of nuclei.[23,24] The most accurate approach to extract nucleation rates from experimental data is by use of the population balance model (PBM)[25]

$$\frac{\partial n}{\partial t} + \frac{\partial(nG_r)}{\partial r} + \frac{\partial(nG_L)}{\partial L} = n\frac{d\ln V_d}{dt} + J\delta(r - R_0)\delta(L - R_0) \quad (1)$$

where $n(t, r, L)$ is the crystal population density (#/μm³μm²), $L$ is a crystal length, $r$ is the crystal width, $R_0$ is the dimension of a crystal nucleus (μm), $t$ is time (h), and $V_d$ is the volume of the droplet (μm³), $J$ is the nucleation rate (#/μm³s), and $G_L$ and $G_r$ are the growth rates (μm/h) along crystal length and radial directions. The initial condition for the PBM for each droplet is $n(0, r, L) = 0$. The PBM is accompanied with the solute mass balance

$$\frac{dc_{rAAV}}{dt} = c_{rAAV}\frac{d\ln V_d}{dt} - J - \frac{1}{v_{m,rAAV}}\int_0^L\int_0^R (\pi r^2 G_L + 2\pi r G_r L) n\, dr\, dL \quad (2)$$

where $c_{rAAV}$ is the capsid concentration in aqueous solution (number of capsids/μm³), which has the initial condition that $c_{rAAV}(0) = c_{rAAV,0}$, and $v_{m,rAAV}$ is the volume of a single rAAV capsid (4.2 × $10^{-6}$ μm³).[26]

**Modeling of droplet volume evolution:** As per thermodynamics, a water vapor pressure depression develops when a salt is added into a solution (colligative property).[27] Since the salt concentration is two-fold higher in the reservoir solution than in the droplet, vapor pressure depression will be more for the reservoir solvent than in the droplet. Thus, vapor diffusion resulting in net transfer of water from the droplet to the reservoir solution occurs. Initially, vapor from both droplet and reservoir diffuses into the open space inside the chamber. Once the droplet water vapor pressure equals the water partial pressure in the air, transfer of the vapor from the droplet takes place to the reservoir water as there will be a net vapor pressure driving force. With the reasonable approximation that the water density is constant through the experiments, the decrease in water volume for the droplet can be written as

$$\frac{dV_d}{dt} = \begin{cases} \dfrac{A_d k\left(p_{vd}(t) - p_{w,\text{air}}(t)\right)}{\rho_w} & \text{for } p_{vd} > p_{w,\text{air}} \quad (3) \\ \dfrac{A_d k(p_{vd}(t) - p_{vR}(t))}{\rho_w} & \text{for } p_{vd} = p_{w,\text{air}} \quad (4) \\ & \text{and } p_{vd} > p_{vR} \end{cases}$$

$$p_{vd}(t) = (1 - x_d(t))p_v^0 \quad (5a)$$
$$p_{vR}(t) = (1 - x_R(t))p_v^0 \quad (5b)$$

$$\frac{dp_{w,\text{air}(t)}}{dt} = \left(\frac{A_d k\left(p_{vd}(t) - p_{w,\text{air}}(t)\right) + A_R k\left(p_{vR}(t) - p_{w,\text{air}}(t)\right)}{M_{t,\text{air}}}\right) p_{t,\text{air}}$$

$$\text{for } p_{vd} > p_{w,\text{air}} \quad (5c)$$

where $p_{vd}$ is the vapor pressure of droplet water, $p_{vR}$ is the vapor pressure of reservoir water, $p_{w,\text{air}}$ is the partial pressure of the water vapor in the air in the open space in the closed well, $p_{t,\text{air}}$ is the total pressure of air, $\rho_w$ is the density of water, $A_d$ is the area of the droplet-air interface, $k$ is the mass transfer coefficient of water between the droplet and the air given in the literature for room temperature and still air,[28–30] $p_v^0$ is the pure water vapor pressure at temperature $T$, $M_{t,\text{air}}$ is the total mass of the air including water vapor, and $x_d$ and $x_R$ are the mole fractions of NaCl in droplet and reservoir solution, respectively. The initial value of the droplet volume is $V_d(0) = V_0$.

**Modeling of nucleation and growth kinetics:** The nucleation rate was modelled by the classical nucleation expression[31]

$$J(t) = k_n \exp\left(-\frac{f\Delta G}{k_B T}\right),$$

$$\Delta G = \frac{16\pi\gamma^3 v_{m,rAAV}^2}{(k_B T)^3 \left(\ln \frac{c_{rAAV}}{c_s}\right)^2}, \quad (6)$$

where $\Delta G$ is the Gibbs free energy change of nuclei formation, $c_s$ is the solubility of rAAV capsids, $\gamma$ is the interfacial tension, $k_B$ is the Boltzmann constant, $T$ is the temperature, $k_n$ is a nucleation pre-factor, and $f$ is the correction in the Gibbs free energy barrier for primary nucleation. As the capsid nucleation mechanism is not well understood, a correction parameter is incorporated in the Gibbs free energy term in the classical nucleation model, where $f = 1$ for homogeneous nucleation and $f < 1$ represents a heterogeneous nucleation mechanism.[31,32] The growth rate models are the widely used power-law expressions[33,34]

$$G_r(t) = k_r [c_{rAAV}(t) - c_s]^{g_r} \quad (7a)$$
$$G_L(t) = k_L [c_{rAAV}(t) - c_s]^{g_L} \quad (7b)$$

where $k_r$ and $k_L$ are growth pre-factors, and $g_r$ and $g_L$ are growth rate exponents (between 1 and 2 for most active pharmaceutical ingredients, APIs).[35–37]

The unknown kinetic parameters $k_n, f, \gamma, k_r, g_r, k_L, g_L$ were fit to direct measurements of $G_r, G_L, n(t,r,L)$, $n_t, L_t, r_t, A_t, V_t$ for multiple droplets, where $n_t, L_t, r_t, A_t, V_t$ are the total number density, total length, total radius, total surface area, and total volume, respectively. The model values for the latter variables are determined by applying the method of moments to the PBM to give[37,38]

$$\frac{dn_t}{dt} = J \quad (8)$$

$$\frac{dL_t}{dt} = JR_0 + \int_0^L \int_0^R n G_L \, dr dL \quad (9)$$

$$\frac{dr_t}{dt} = JR_0 + \int_0^L \int_0^R n G_r \, dr dL \quad (10)$$

$$\frac{dA_t}{dt} = JA_n + 2\pi \int_0^L \int_0^R n(LG_r + rG_L + 2rG_r) \, dr dL \quad (11)$$

$$\frac{dV_t}{dt} = Jv_n + \pi \int_0^L \int_0^R n(2rG_r L + r^2 G_L) \, dr dL \quad (12)$$

where $A_n$ and $v_n$ are the nuclei surface area and volume, respectively. More specifically, the parameters were fit by maximum likelihood estimation, which solves the optimization

$$\min_{k'} \left(y_e - y_s(k', k_0)\right)^T V_e^{-1}(k')\left(y_e - y_s(k', k_0)\right) \quad (13)$$

and

$$V_e = \left.\frac{\partial y_s}{\partial k_0}\right|_{k_0} V_{k_0} \left.\frac{\partial y_s}{\partial k_0}\right|_{k_0}^T + V_\varepsilon \quad (14)$$

where $k'$ is the vector of parameter estimates, $k_0$ is the vector of predetermined parameters, $y_e$ is the vector of experimental observations, $y_s$ is the vector of model predictions, and $V_\epsilon$ is the matrix of experimental data variance, which considers the effect of predetermined parameter variances, and normal noise $e$. Detailed derivations can be found in prior publications.[36,37,39]

Simulation and optimization are carried out in MATLAB. The partial differential equation (1) is discretized in the size domain using finite difference scheme to obtain a set of ordinarily differential equations (ODEs) in the time domain. This set of ODEs is solved using a stiff differential equation solver ode15s in MATLAB. The parameters are fit using a multi-start optimization approach with sequential quadratic programming (SQP) used to search for a global minimum by minimizing objective function locally from 1100 starting points.

### 3. Results and Discussion

Control experiments and solubility analysis were carried out to rule out the possibility that any of the particles formed in the experiment consisted of NaCl or PEG.[40] Crystals were observed under cross-polarized light microscope to confirm the crystallinity of particles. For detailed description of polarized light microscopic identification of crystals, readers are referred to the previous article from this group.[40] Cross-polarized light microscopic images of crystals in a droplet at different magnifications are shown in **Fig. 3**. Crystals are mostly cylindrical/rod/flat bar in shape.

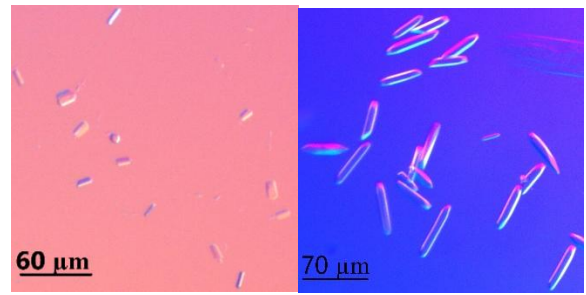

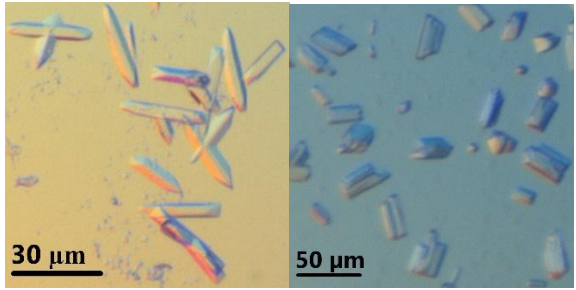

**Fig. 3:** Images of crystals under cross polarizer microscope. Row 1: (left) "empty" (2% PEG, 0.6 M NaCl, and 5.7 pH) and (right) "full" (3% PEG, 1 M NaCl, and 5.7 pH) capsids crystals of rAAV5. Row 2: (left) "empty" (0.3% PEG, 3 M NaCl, and 5.7 pH) and (right) "full" (3.5% PEG, 0.4 M NaCl, and 5.7 pH) capsids crystals of rAAV8.

The population balance equation with classical exponential nucleation and power law growth models[33,34] agree with the experimental data for the crystal size distribution and crystal growth rates for full rAAV5 and full rAAV8 (**Fig. 4ab** & **5**). The model results also agree with the experimental data for the total number density variation, total length of crystals, total radius of crystals, total crystal surface area, and total volume of the crystals with RMSE < 0.21 (**Figs. 6–10**, respectively). The power law nucleation expression predicts that nucleation occurs for positive supersaturation no matter how small its value. In reality, nucleation does not occur for a long duration of time (**Fig. 6**). The fall in nucleation rate as the supersaturation drops is captured well by the classical exponential nucleation model (eq. 6) (**Fig. 11**).

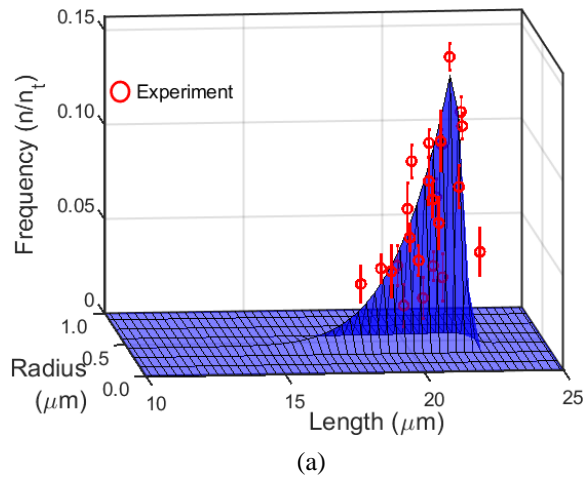

(a)

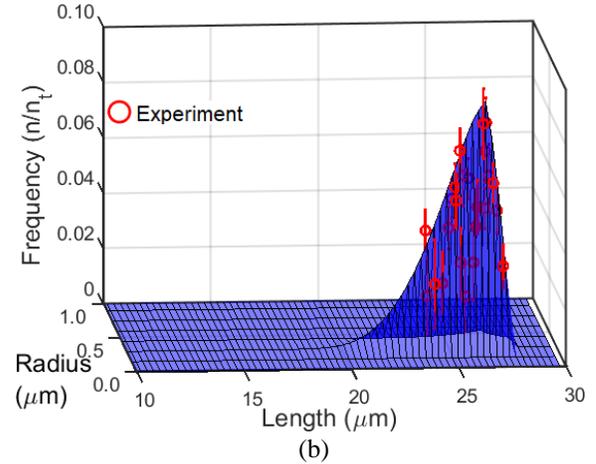

(b)

**Fig. 4:** Comparison of model results with experimental data for crystal size distribution for (a) full rAAV5 and (b) rAAV8 at time $t = 60$ h, 2.5% PEG, 0.75 M NaCl, and pH 6 (see Table 1 for parameter estimates). Solubility: $5 \times 10^{10}$ vg/µL for rAAV5 and $4.6 \times 10^{10}$ vg/µL for rAAV8.

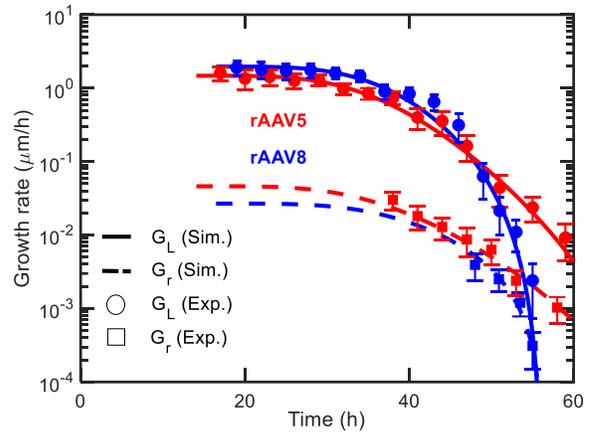

**Fig. 5:** Comparison of simulation results with experimental data for growth rate variation for full rAAV5 and rAAV8 at 2.5% PEG, 0.75 M NaCl, and pH 6 (see Table 1 for parameter estimates). Solubility: $5 \times 10^{10}$ vg/µL for rAAV5 and $4.6 \times 10^{10}$ vg/µL for rAAV8.

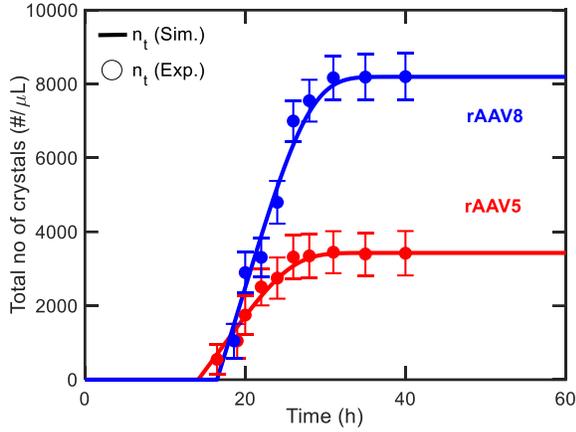

**Fig. 6:** Comparison of model results with experimental data for total number density of crystals for full rAAV5 and rAAV8 at 2.5% PEG, 0.75 M NaCl, and pH 6 (see Table 1 for parameter estimates). Solubility: $5 \times 10^{10}$ vg/µL for rAAV5 and $4.6 \times 10^{10}$ vg/µL for rAAV8.

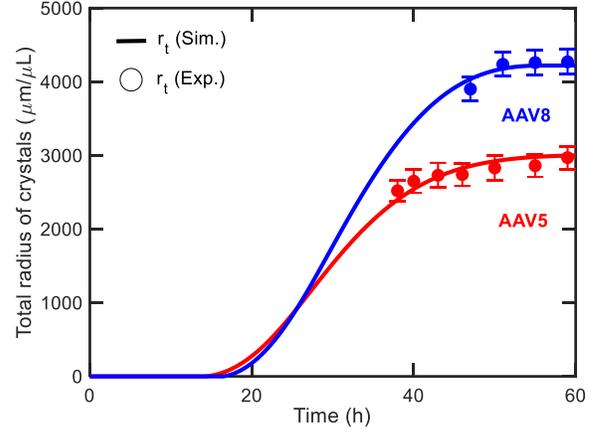

**Fig. 8:** Comparison of model simulation results with that of experimental data for total radius of crystals for full rAAV5 and rAAV8 at 2.5% PEG, 0.75 M NaCl, and pH 6 (see Table 1 for parameter estimates). Solubility: $5 \times 10^{10}$ vg/µL for rAAV5 and $4.6 \times 10^{10}$ vg/µL for rAAV8.

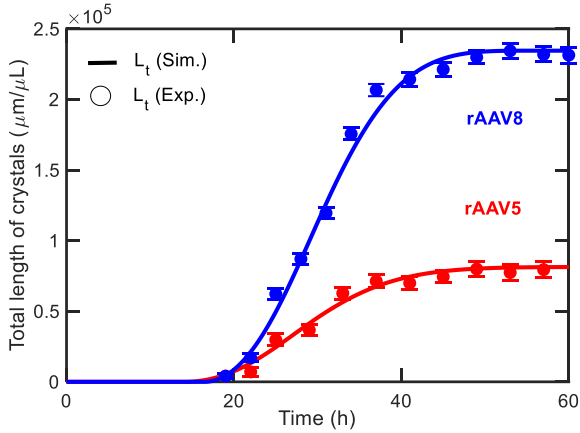

**Fig. 7:** Comparison of model simulation results with that of experimental data for total length of crystals for full rAAV5 and rAAV8 at 2.5% PEG, 0.75 M NaCl, and pH 6 (see Table 1 for parameter estimates). Solubility: $5 \times 10^{10}$ vg/µL for rAAV5 and $4.6 \times 10^{10}$ vg/µL for rAAV8.

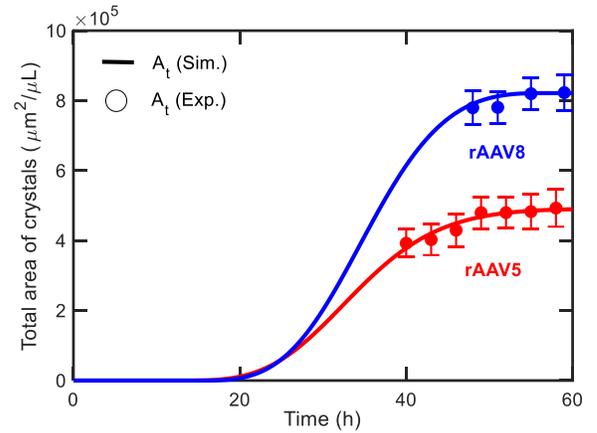

**Fig. 9:** Comparison of model simulation results with that of experimental data for total surface area of crystals for full rAAV5 and rAAV8 at 2.5% PEG, 0.75 M NaCl, and pH 6. (see Table 1 for parameter estimates). Solubility: $5 \times 10^{10}$ vg/µL for rAAV5 and $4.6 \times 10^{10}$ vg/µL for rAAV8.

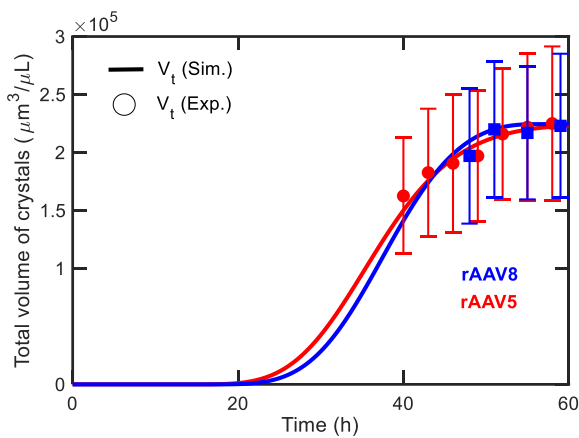

**Fig. 10:** Comparison of model results with experimental data for total volume of crystals for full rAAV5 and rAAV8 at 2.5% PEG, 0.75 M NaCl, and pH 6 (see Table 1 for parameter estimates). The larger relevative error bars compared to other measurements are due to the errors associated with radius and length measurement at short length scales. Solubility: $5 \times 10^{10}$ vg/μL for rAAV5 and $4.6 \times 10^{10}$ vg/μL for rAAV8.

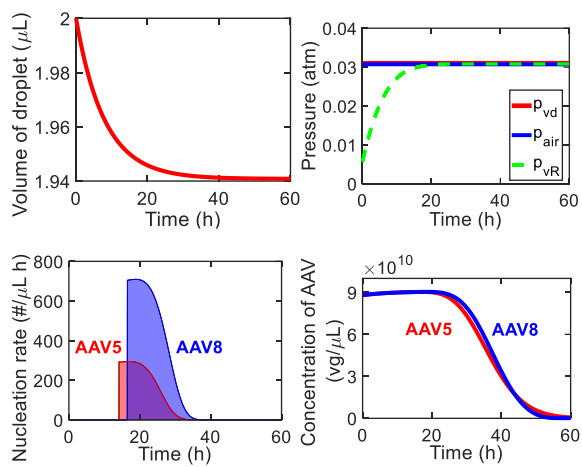

**Fig. 11:** Model results for droplet volume variation (top left), vapor pressure and air pressure variation (top right), full rAAV nucleation rate variation (bottom left), and concentration variation (bottom right) at 2.5% PEG, 0.75 M NaCl, and pH 6 (see Table 1 for parameter estimates).

For both full rAAV5 and full rAAV9, the first crystal nuclei appear after a long period 13 - 16 h (**Figs. 6** and **11**). Vapor diffusion causes the droplet to shrink by ~ 2.5% in volume ~ 13 h after the onset of experiment (**Fig. 11**) and this causes the the rAAV concentration to increase by about 2.5%. The rAAV concentration reaches a maximum at ~ 20 h (**Fig. 11**). Although thermodynamics suggests that the droplet volume should shrink by 50%, the kinetics of vapor diffusion from the droplet to the reservoir becomes so slow after a day that the droplet volume is reduced by only 3% even after a couple of weeks in contrast to that reported in the literature.[41]

The model indicates that the nucleation induction time, which is the time period between the instant when the suoersaturation is generated and the time instant when the particles are formed, can range between 14 to 38 h (**Table 1**), depending on the precipitant concentrations. This induction time scales are quite similar to those (10 to 25 h) obtained for small-molecule APIs such as paracetamol in hanging drop evaporation-based microfluidic system of starting volume 5 μL.[14,15] The calculated induction time scales are somewhat higher than that reported for glycine for a droplet evaporation-based microfluidic system of a starting droplet volume of 4.5 μL.[14] The comparable induction times are in constrast to the orders of magnitude larger molecular weight of capsids compared to glycine or paracetamol. In addition, capsid nucleation in hanging-drop vapor diffusion system occurs at low supersaturation (~1.62) compared to the nucleation of glycine or paracetamol (supersaturation > 4) in a droplet evaporation-based microfluidic system.[14,15,42] As such, the capsid nucleation rate/propensity is higher for capsids than for small molecules such as paracetamol and glycine. This observation conflicts with the literature reports that the propensity of proteins/macromolecules to nucleate decreases as the molecular weight increases, due to the increase in the number of degrees of freedom/conformations making the conformational matching with the lattice sites increasingly difficult.[41] Capsids have an icosahedral symmetry (T=1) and 2-fold, 3-fold, and 5-fold symmetry with a surface decorated with recurrent protrusions of various sizes.[19] This ball-shaped outer structure with low degrees of freedom facilitates conformational matching required to sit on the lattice sites to form a crystal, i.e., the overall symmetry and physically constrained capsids minimize the orientation degrees of freedom to achieve crystal formation. Thus capsids possess higher propensity to nucleate compared to many small molecules, macromolecules, and heavy proteins even though capsids possess a molecular weight (3.6 MDa) that is much higher than many of the heavy macromolecules such as mAbs (~150 kDa) and mRNAs (~200–500 kDa).[1,43]

The difference in induction time scale between two serotypes is not significant given the concnetrations of the precipitants used in the simulation are very similar for both. For example, for precipitant concentrations of 2.5 w/v% PEG and 0.75 M NaCl, the induction time scales for full rAAV5 and full rAAV8 are 14.1 h and 16.5 h, respectively, with only a ~2 h time difference (**Table 1**). Considering the stochastic nature of the nucleation event, particularly in the droplet-based system of microliter volume, this difference in time scale may not be statistically significant as the induction time in these systems may range more than 10 h.[44]

In both the serotypes, nucleation lasts for a period of ~10–24 h (**Fig. 11**). The prolonged nucleation is associated with a broad size distribution for both serotypes (**Figs. 4ab**). Capsid nucleation is a slow process compared to the LaMer burst nucleation (i.e., instant nucleation resulting in monodisperse particles) observed in chalcogenides/metals/metal-oxides/some semiconductors, but this prolonged nucleation period is comparable to that observed in the crystallization of lysozyme, a low-molecular-weight protein, in a continuous stirred tank reactor of 100 mL.[13,45–47] However, the nucleation rate starts decreasing as soon as the solute concentration begins to drop (4–5 h after the onset, see **Fig. 11**) and eventually nucleation stops. The classical nucleation model captures this duration quite well with a dimensionless correction parameter $f$ of $7.9 \times 10^{-7}$ and $7.94 \times 10^{-7}$ for full rAAV5 and full rAAV8, respectively and a nucleation pre-factor $k_n$ of $\sim 1 \times 10^4$ #/µm$^3$s and $\sim 1 \times 10^3$ #/µm$^3$s for full rAAV5 and full rAAV8, respectively. The value of the nucleation pre-factor is found to be lower than that of $10^7$–$10^{17}$ #/µm$^3$s for primary homogeneous nucleation of metal and oxides reported in literature.[48]

The extremely low values of free energy correction factor indicate that the free energy barrier for nucleation is quite low. As such, the nucleation is feasible at low supersaturation which suggests that the nucleation process is dominated by a heterogeneous nucleation mechanism. A ~2.5% increase in supersaturation resulting from ~2.5% decrease in droplet volume due to vapor diffusion was sufficient to cause the onset of nucleation as the free energy barrier for heterogeneous nucleation is relatively low. Though nucleation periods and induction times for capsids and low-molecular-weight proteins such as lysozyme are comparable, the growth rate prefactors for the former are almost seven orders of magnitude smaller than that of the latter and the growth rate exponent is nearly half of that of the latter. The growth of capsid crystals could be limited by the transport of the capsids to the crystal surface, surface integration, or 2D surface diffusion. Integration into a kink site is unlikely to be limiting since the capsids have few orientation degrees of freedom. Having one of the transport steps being limiting is probably characteristic of macromolecular assemblies such as capsids possessing extremely high mass (resulting in slow Brownian motion), which prevents the capsids from approaching to crystal surface to integrate into the crystal or limits the surface diffusion. This effect of high molecular mass is incorporated implicitly into the growth rate prefactor and reflected by its very low value.

The aspect ratio (length/diameter) of full rAAV5 crystals is smaller than for full rAAV8 crystals (**Figs. 4ab**), due to the faster axial growth and slower radial growth for full rAAV8 compared to full rAAV5 (**Fig. 5**). rAAV8 crystals have higher total length (**Fig. 7**), total radius (**Fig. 8**), and total surface area (**Fig. 9**) than rAAV5. The lower aspect ratio of the rAAV5 crystals is associated with the lower total surface area and a lower surface area-to-volume ratio (cf., **Figs. 9 & 10**). While the full rAAV8 crystals have a higher total number density (**Fig. 6**), the full rAAV5 crystals have higher growth rates (**Fig. 5**) so that average mass per crystal is higher, and the total volume of crystals is the same for both serotypes throughout the course of the experiments (**Fig. 10**).

The nucleation and growth kinetics and the time scale for onset of nucleation were calculated for different experimental conditions (**Table 1**). As expected, the time scale for onset of nucleation varies nonlinearly with the concentration of the precipitant. The data also show the strongly nonlinear interaction between precipitant, solute, and solvent. Variation in the values of nucleation and growth kinetics from different sets of experiments probably suggests the dependence of nucleation rate on many properties such as surface tension variation with precipitant, solute-solvent-precipitant interaction, and variation in cluster-cluster collision frequency.

**Table 1:** Values of nucleation and growth kinetic constants for experiments at a pH 6 and capsids at $8 \times 10^{10}$ vg/μL.

| | PEG w/v % | NaCl (M) | $t_n$ (h) | $L_c$ (μm) | $k_0 \times 10^{-3}$ [#(vg)(μm)$^{-3}$s$^{-1}$] | $f \times 10^7$ | $k_L \times 10^6$ [(μm)$^{(3g_L+1)}$vg$^{-g_L}$s$^{-1}$] | $g_L$ | $k_r \times 10^7$ [(μm)$^{(3g_r+1)}$vg$^{-g_r}$s$^{-1}$] | $g_r$ |
|---|---|---|---|---|---|---|---|---|---|---|
| rAAV5 full | 2.50 | 0.75 | 14.1 | 22.0 | 10.00 | 7.900 | 2.90 | 1.10 | 3.50 | 0.80 |
| | 2.60 | 0.83 | 17.2 | 41.0 | 15.30 | 7.910 | 2.79 | 1.09 | 3.56 | 0.82 |
| | 2.60 | 0.92 | 18.3 | 35.5 | 13.35 | 7.902 | 2.85 | 1.07 | 3.51 | 0.87 |
| | 2.70 | 1.27 | 21.5 | 37.2 | 17.85 | 7.905 | 2.87 | 1.12 | 3.48 | 0.78 |
| rAAV5 empty | 1.98 | 0.15 | 36.7 | 39.7 | 12.40 | 7.897 | 2.96 | 1.11 | 3.45 | 0.81 |
| | 2.00 | 0.20 | 32.5 | 35.1 | 13.70 | 7.908 | 2.98 | 1.09 | 3.57 | 0.85 |
| | 2.30 | 0.27 | 29.2 | 43.0 | 14.18 | 7.898 | 3.06 | 1.11 | 3.53 | 0.89 |
| rAAV8 full | 2.50 | 0.75 | 16.5 | 26.0 | 1.00 | 7.940 | 3.85 | 1.10 | 5.00 | 0.60 |
| | 2.70 | 0.05 | 22.3 | 24.0 | 1.30 | 7.937 | 3.95 | 1.04 | 5.07 | 0.57 |
| | 2.75 | 0.10 | 24.5 | 25.4 | 1.45 | 7.936 | 3.89 | 1.25 | 5.11 | 0.62 |
| | 2.75 | 0.20 | 26.1 | 29.9 | 2.13 | 7.945 | 3.84 | 1.01 | 5.16 | 0.65 |
| rAAV8 empty | 3.10 | 0.06 | 31.3 | 28.5 | 2.51 | 7.942 | 3.92 | 1.17 | 4.96 | 0.56 |
| | 3.10 | 0.15 | 38.4 | 19.3 | 3.07 | 7.946 | 3.86 | 1.13 | 4.97 | 0.53 |
| | 3.20 | 0.15 | 36.1 | 18.5 | 3.12 | 7.939 | 3.88 | 1.09 | 4.93 | 0.57 |

Key: $t_n$: induction time/nucleation onset time scale, $L_c$: crystal length

## 5. Conclusion

We derived a mechanistic model for AAV capsid crystallization based on coupling population balance and species balance equations which (1) includes the effect of droplet volume evolution and solution thermodynamics and (2) captures the experimental variation of crystallization kinetics and crystal properties such as length, radius, surface area, and volume with inclusion of a correction factor in the classical exponential nucleation and power law growth models. For the first time, it is found that the crystal nucleation of AAV capsids in a microliter-volume hanging-droplet occurs by heterogeneous nucleation and the nucleation process is driven by water molecule diffusion dynamics from the droplet. The vapor diffusion from the droplet to the reservoir eventually is so slow that the droplet volume shrinks by only 2.5% after a couple of weeks, in contrast to the theoretical equilibrium value of 50% reduction in droplet volume, which is similar to values reported for protein crystallization in droplet-based systems.[41] Heterogeneous nucleation makes crystallization feasible at the low supersaturation associated with the 3% droplet volume shrinkage. Nucleation induction times for capsids are comparable to that of small moelcules such as glycine and paracetamol (API). Despite the capsid's much higher molecular weight, they possess similar propensity to nucleate as low-molecular-weight proteins and small-molecule pharmaceuticals (APIs). The capsid's icosahedral symmetry and 2-fold, 3-fold, and 5-fold symmetry with ball-shaped outer shape/structure,[19] which is unique among macromolecules/macromolecular assemblies/heavy proteins, has low conformational degrees of freedom which facilitates faster alignment into an ordered arrangement to form a crystal nucleus. Thus capsids show higher tendency to form nuclei and crystallize than the other macromolecules such as mAbs or mRNAs.

As with other proteins/biological macromolecules, capsids show a prolonged nucleation period, which leads to the formation of polydispersed crystals. Capsid crystals have a slow growth rate, with growth rate prefactors seven orders of magnitude smaller than that of lysozyme and growth rate exponent nearly half of that of lysozyme, and the crystal growth rate for capsids is more weakly sensitive to the supersaturation than for lysozyme. Slow growth rate is probably limited by either the transport of capsids to the crystal surface or 2D surface transport, because of the slow Brownian motion of larger macromolecular assemblies such as capsids. Finally, the modelling framework presented herein can be useful to explain the crystallization process and predict and control crystallization conditions of other advanced therapeutics/biological micromolecules such as mRNA and mAbs as well.

**Author contributions:**

V.B. conceptualized this work, designed and conducted the experiments, developed models, wrote code, performed simulation, analyzed the data, and wrote the initial draft. M.S.H. designed the automated hanging-drop vapor diffusion system and edited the manuscript. R.D.B. conceptualized this work, supervised the work/project, edited the manuscript, and acquired the funds to support the project. R.M.K., P.W.B., and S.L.S. edited the manuscript, supervised the work/project and acquired the funds to support the project. A.J.S. and J.M.W. edited the manuscript, supervised the project and acquired the funds to support the project.

**Acknowledgements:**
Funding is acknowledged from the MLSC, Sanofi, Sartorius, Artemis, and USFDA (75F40121C00131).

**References:**

(1) Trilisky, E.; Gillespie, R.; Osslund, T. D.; Vunnum, S. Crystallization and Liquid-Liquid Phase Separation of Monoclonal Antibodies and Fc-Fusion Proteins: Screening Results. *Biotechnol. Prog.* **2011**, *27*, 1054–1067.

(2) Krauss, I. R.; Merlino, A.; Sica, F. An Overview of Biological Macromolecule Crystallization. *Int. J. Mol. Sci.* **2013**, *14*, 11643–11691.

(3) Yingxin, L.; Wang, X.; Ching, C. B. Toward Further Understanding of Lysozyme Crystallization: Phase Diagram, Protein-Protein Interaction, Nucleation Kinetics, and Growth Kinetics. *Cryst Growth Des* **2010**, *10*, 548–558.

(4) Bessho, Y.; Ataka, M.; Asai, M.; Katsura, T. Analysis of the Crystallization Kinetics of Lysozyme Using a Model with Polynuclear Growth Mechanism. *Biophys J* **1994**, *66*, 310–313.

(5) Lerch, T. F.; Xie, Q.; Ongley, H. M.; Hare, J.; Chapman, M. S. Twinned Crystals of Adeno-Associated Virus Serotype 3b Prove Suitable for Structural Studies. *Acta Crystallogr Sect F Struct Biol Cryst Commun* **2009**, *65*, 177–183.

(6) Xie, Q.; Bu, W.; Bhatia, S.; Hare, J.; Somasundaram, T.; Azzi, A.; Chapmanm, M. S. The Atomic Structure of Adeno-Associated Virus (AAV-2), a Vector for Human Gene Therapy. *Proc Natl Acad Sci USA* **2002**, *99*, 10405–10410.

(7) Hong, M. S.; Lu, A. E.; Bae, J.; Lee, J. M.; Braatz, R. D. Droplet-Based Evaporative System for the Estimation of Protein Crystallization Kinetics. *Cryst. Growth Des.* **2021**, *21*, 6064–6075.

(8) Miller, E. B.; Whitaker, B. G.; Govindasamy, L.; McKenna, R.; Zolotukhin, S.; Muzyczka, N.; McKenna, M. A. Production, Purification and Preliminary X-Ray Crystallographic Studies of Adeno-Associated Virus Serotype 1. *Acta Crystallogr Sect F Struct Biol Cryst Commun* **2006**, *62*, 1271–1274.

(9) Xie, Q.; Ongley, H. M.; Hare, J.; Chapman, M. S. Crystallization and Preliminary X-Ray Structural Studies of Adeno-Associated Virus Serotype 6. Acta Crystallogr Sect F Struct Biol Cryst Commu. *Acta Crystallogr Sect F Struct Biol Cryst Commun* **2008**, *64*, 1074–1078.

(10) Zhang, B.; Wang, Y.; Thi, S.; Toong, V.; Luo, P.; Fan, S.; Xu, L.; Yang, Z.; Heng, J. Y. Y. Enhancement of Lysozyme Crystallization Using DNA as a Polymeric Additive. *Crystals (Basel)* **2019**, *9*, 186.

(11) Pal, K.; Yang, Y.; Nagy, Z. K. Model-Based Optimization of Cooling Crystallization of Active Pharmaceutical Ingredients Undergoing Thermal Degradation. *Cryst. Growth Des.* **2019**, *19*, 3417–3429.

(12) Variankaval, N.; Cote, A. S.; Doherty, M. F. From Form to Function: Crystallization of Active Pharmaceutical Ingredients. *AIChE Journal* **2008**, *54*, 1682–1688.

(13) Rosenbaum, T.; Tan, L.; Engstrom, J. Advantages of Utilizing Population Balance Modeling of Crystallization Processes for Particle Size Distribution Prediction of an Active Pharmaceutical Ingredient. *Processes* **2019**, *7*, 355.

(14) Chen, K.; Goh, L.; He, G.; Kenis, P. J. A.; Zukoski, C. F.; Braatz, R. D. Identification of Nucleation Rates in Droplet-Based Microfluidic Systems. *Chem Eng Sci* **2012**, *77*, 235–241.

(15) Goh, L.; Chen, K.; Bhamidi, V.; He, G.; Kee, N. C. S.; Kenis, P. J. A.; Zukoski, C. F.; Braatz, R. D. A Stochastic Model for Nucleation Kinetics Determination in Droplet-Based Microfluidic Systems. *Cryst Growth Des* **2010**, *10*, 2515–2521.

(16) Mitchell, H. M.; Jovannus, D.; Rosbottom, I.; Link, F. J.; Mitchell, N. A.; Heng, J. Y. Y. , Process Modelling of Protein Crystallisation: A Case Study of Lysozyme. *Chemical Engineering Research and Design* **2023**, *192*, 268–279.

(17) Saikumar, M. V.; Glatz, C. E.; Larson, M. A. Lysozyme Crystal Growth and Nucleation Kinetics. *J Cryst Growth* **1998**, *187*, 277–288.

(18) Talreja, S.; Kenis, P. J. A.; Zukoski, C. F. A Kinetic Model To SimulatProtein Crystal Growth in an Evaporation-Based Crystallization Platform. *Langmuir* **2007**, *23*, 4516–452210.

(19) Wörner, T. P.; Bennett, A.; Habka, S.; Snijder, J.; Friese, O.; Powers, T.; Agbandje-McKenna, M.; Heck, A. J. R. Adeno-Associated Virus Capsid Assembly Is Divergent and Stochastic. *Nat Commun* **2021**, *12*, 1642.

(20) Montiel-Garcia, D.; Santoyo-Rivera, N.; Ho, P.; Carrillo-Tripp, M.; Johnson, J. E.; Reddy, V. S. VIPERdb v3. 0: A Structure-Based Data Analytics Platform for Viral Capsids. *Nucleic Acids Res* **2020**, *49*, D809–D816.

(21) Rayaprolu, V.; Kruse, S.; Kant, R.; Kant, R.; Venkatakrishnan, B.; Movahed, N.; Brooke, D.; Lins, B.; Bennett, A.; Potter, T.; McKenna, R.; Agbandje-McKenna, M.; Bothner, B. Comparative Analysis of Adeno-Associated Virus Capsid Stability and Dynamics. *J Virol* **2013**, *87*, 13150–13160.

(22) NIH 3D. *Structure of Adeno-Associated virus serotype 8*. NIH 3D.

(23) McGinty, J., Yazdanpanah, N., Price, C., ter Horst, J., & Sefcik, J. Nucleation and Crystal Growth in Continuous Crystallization. In *The handbook of continuous crystallization*; Yazdanpanah, N., Nagy, Z. K., Ed.; Royal society of chemistry, 2020; pp 1–50.

(24) Strey, R.; Wagner, P. E.; Viisanen, Y. The Problem of Measuring Homogeneous Nucleation Rates and the Molecular Contents of Nuclei: Progress in the Form of Nucleation Pulse Measurements. *Journal of Physical Chemistry* **1994**, *98* (32), 7748–7758. https://doi.org/10.1021/j100083a003.

(25) Ramkrishna, D. *Population Balances: Theory and Applications to Particulate Systems in Engineering*, 1st ed.; Academic Press, 2000.

(26) Snyder, R.O., & Moullier, P. *Adeno-Associated Virus: Methods and Portocols*, 1st ed.; Walker, J. M., Ed.; Springer: New York, 2011.

(27) Sandler, S. I. *Chemical, Biochemical, and Engineering Thermodynamics*, 5th ed.; Wiley: Indianapolis, USA, 2017.

(28) Sparrow, E. M.; Kratz, G. K.; Schuerger, M. J. Evaporation of Water from a Horizontal Surface by Natural Convection. *J Heat Transfer* **1983**, *105*, 469–475.

(29) Talev, G.; Thue, J. V.; Gustavsen, A. *Measurements of the convective mass transfer coefficient between the water surfaces and still air*.

(30) Poós, T.; Varju, E. Mass Transfer Coefficient for Water Evaporation by Theoretical and Empirical Correlations. *Int J Heat Mass Transf* **2020**, *153*, 119500.

(31) Fletcher, N. H. Size Effect in Heterogeneous Nucleation. *J. Chem. Phys.* **1958**, *29*, 572–576.

(32) Liu, X. Y. Heterogeneous Nucleation or Homogeneous Nucleation? *J. Chem. Phys.* **2000**, *112*, 9949–9955.

(33) Marchiso, D. L.; Barresi, A. A.; Garbero, M. Nucleation, Growth, and Agglomeration in Barium Sulfate Turbulent Precipitation. *AIChE Journal* **2002**, *48* (9), 2039–2050. https://doi.org/10.1002/aic.690480917.

(34) Woo, X. Y.; Tan, R. B. H.; Chow, P. S.; Braatz, R. D. Simulation of Mixing Effects in Antisolvent Crystallization Using a Coupled CFD-PDF-PBE Approach. *Cryst Growth Des* **2006**, *6* (6), 1291–1303. https://doi.org/10.1021/cg0503090.

(35) Kee, N. C. S.; Arendt, P. D.; Goh, L. M.; Tan, R. B. H.; Braatz, R. D. Nucleation and Growth Kinetics Estimation for L-Phenylalanine Hydrate and Anhydrate Crystallization. *Cryst. Eng. Comm.* **2011**, *13*, 1197–2009.

(36) Nagy, Z. K.; Fujiwara, M.; Woo, X. Y.; Braatz, R. D. Determination of the Kinetic Parameters for the Crystallization of Paracetamol from Water Using Metastable Zone Width Experiments. *Ind Eng Chem Res* **2008**, *47* (4), 1245–1252. https://doi.org/10.1021/ie060637c.

(37) Gunawan, Rudiyanto.; Ma, D. L.; Fujiwara, M.; Braatz, R. D. Identification of Kinetic Parameters in Multidimensional Crystallization Processes. *International journal of modern physics* **2002**, *16*, 367–374.

(38) Nagy, Z. K.; Fujiwara, M.; Woo, X. Y.; Braatz, R. D. Determination of the Kinetic Parameters for the Crystallization of Paracetamol from Water Using Metastable Zone Width Experiments. *Ind. Eng. Chem. Res.* **2008**, *47*, 1245–1252.

(39) Nguyen, T. N. T.; Sha, S.; Hong, M. S.; Maloney, A. J.; Barone, P. W.; Neufeld, C.; Wolfrum, J.; Springs, S. L.; Sinskey, A. J.; Braatz, R. D. Mechanistic Model for Production of Recombinant Adeno-Associated Virus via Triple Transfection of HEK293 Cells. *Mol Ther Methods Clin Dev* **2021**, *21*, 642–655. https://doi.org/10.1016/j.omtm.2021.04.006.

(40) Bal, V.; Wolfrum, J. M.; Barone, P. W.; Springs, S. L.; Sinskey, A. J.; Kotin, R. M.; Braatz, R. D. *Selective Enrichment of Full AAV Capsids*; 2024. https://doi.org/arXiv:2412.06093.

(41) McPherson, A.; Gavira, J. A. Introduction to Protein Crystallization. *Acta Crystallographica Section F:Structural Biology Communications* **2014**, *70*, 2–20.

(42) Talreja, S.; Kim, D. Y.; Mirarefi, A. Y.; Zukoskia, C. F.; Kenis, P. J. A. Screening and Optimization of Protein Crystallization Conditions through Gradual Evaporation Using a Novel Crystallization Platform. *J. Appl. Cryst.* **2005**, *38*, 988–995.

(43) Rakel, N.; Baum, M.; Hubbuch, J. Moving through Three-Dimensional Phase Diagrams of Monoclonal Antibodies. *Biotechnol. Prog.* **2014**, *13*, 1103.

(44) Goh, L.; Chen, K.; Bhamidi, V.; He, G.; Kee, N. C. S.; Kenis, P. J. A.; Zukoski, C. F.; Braatz, R. D. A Stochastic Model for Nucleation Kinetics Determination in Droplet-Based Microfluidic Systems. *Cryst Growth Des* **2010**, *10*, 2515–2521.

(45) Kwon, S. G. A. T. H.; Hyeon, T. Colloidal Chemical Synthesis and Formation Kinetics of Uniformly Sized Nanocrystals of Metals, Oxides, and Chalcogenides. *ACCOUNTS OF CHEMICAL RESEARCH* **2008**, *41*, 1696–1709.

(46) Whitehead, C. B.; zkar, S. O.; Finke, R. G. LaMer's 1950 Model of Particle Formation: A Review and Critical Analysis of Its Classical Nucleation and Fluctuation Theory Basis, of Competing Models and Mechanisms for Phasechanges and Particle Formation, and Then of Its Application to Silver Halide, Semiconductor, Metal, and Metal-Oxide Nanoparticles. *Mater. Adv.* **2021**, *2*, 186.

(47) Tian, W.; Li, W.; Yang, H. Protein Nucleation and Crystallization Process with Process Analytical Technologies in a Batch Crystallizer. *Cryst. Growth Des.* **2023**, *23*, 5181–5193.

(48) Randolph, A. D.; Larson, M. A. *Theory of Particulate Processes*; Elsevier: Amsterdam, 1971.